\documentclass[aps,prd,reprint,nofootinbib]{revtex4-2}
\usepackage{graphicx}
\usepackage{amssymb}
\usepackage{amsmath}
\usepackage{lipsum}

\usepackage{etoolbox}
\apptocmd{\thebibliography}{\raggedright}{}{}

\begin{document}

\title{An interpretation of scalars in SO(32)}

\author{Alejandro Rivero}
\affiliation{Institute for Biocomputation and Physics of Complex Systems (BIFI), Universidad de Zaragoza, Spain}
\begin{abstract}
We propose an interpretation for the adjoint representation
of the $SO(32)$ group to classify the
scalars of a generic Supersymmetric Standard Model having just three
generations of particles, via a flavour group $SU(5)$. We show that
this same interpretation arises from a simple postulate of 
self-consistence of composites for these scalars. The model 
looks only for colour and electric charge, and it pays the
cost of an additional chiral $+4/3$ quark per generation.
\end{abstract}
\maketitle

 






\section{Introduction}
\label{introduction}

While highly relevant in string theory and supergravity, the $SO(32)$ group is not a good unification group as it
doesn't have complex representations \citep{Gell-Mann:1979vob}. But it still gets an interesting family group when decomposed.
In this letter, we first review the decomposition, interpret it as a group symmetry on scalars that
could be supersymmetry partners of the Standard Model fermions, and then we present an interesting reconstruction of such scalars as composites. 
Besides, the interpretation has a uniqueness that limits the number of generations for the SM group. 

This reconstruction could have some application when
considering open string theory and their branes, or could be used as basis for other GUT-flavour models. Considering
this, we include a pair of sections with 
some separate discussion on other related groups.

\section{The flavour group in $SO(32)$}
\label{flavourGroup}
The authors of \citep{Gell-Mann:1976xin} classify decomposition of groups
having explicitly a $SU(3)$ colour subgroup, giving candidate
representations as well as the decomposition of the adjoint representation in all the cases.
Groups $SO(2n)$ are case 4 of this classification, where they obtain the decomposition
$SO(n_1)\otimes SU(n_2)\otimes SU(3) \otimes U_1(1)$
with $2n=n_1+6n_2$. Our case of interest is $SO(32)$ with the 
maximal $SU(n_2)$, this is $n_2=5$. The representations
intended for fermions are not very useful, as the group is of
kind $SO(4k)$, without complex representations. But 
we are interested on the adjoint as a place for scalars.
The stated result gives us 
\begin{equation}
\label{original}
\begin{array}{llll}
496=\\
{\bf (1,24,1^c) }&+{\bf [1,15,\bar 3^c]}&+{\bf [1, \bar {15}, 3^c]}&+\\
1,24,8^c&+[1,10,\bar 6^c]&+[1,\bar {10},6^c]&+\\
(1,1,8^c)&&&+\\&(2,5,3^c)&+(2,\bar 5,\bar 3^c)&+\\
&(1,1,1^c)&+[1,1,1^c]\\
\end{array}
\end{equation}
And our components of interest are the three first ones, that we have
stressed with boldface. 
The explicit $U_1(1)$ group provides an hypercharge that counts
the number of coloured representations and is zero
for colour singlets, so we can assign respectively $Y_1=0, +1, -1$ to
the above $1^c$, $\bar 3^c$ and $\bar 3$.

To get a second hypercharge, we can consider $SU(5)$ as the flavour group 
and decompose it \cite{Yamatsu:2015npn} down to multiplets in $SU(3) \times SU(2) \times U_2(1)$
\begin{eqnarray}
15 =& (1, 3)_{-6} + (3, 2)_{-1} + (6, 1)_4 \label{15}  \\
24 =& (1, 1)_0 + (1, 3)_0 + (3, 2)_5 + (\bar 3, 2)_{-5} + (8, 1)_0
\label{24}
\end{eqnarray}
 Now from the two hypercharges we can produce a charge 
 \begin{equation}
 Q= \frac15   \big(\frac 23Y_1 - Y_2\big)
 \end{equation}
and check that the resulting decomposition includes content
corresponding to the scalars of a minimal, three generations, supersymmetric standard model.

\begin{equation}
\begin{array}{lcrrr}
& (f_3, f_2)& Y_1 & Y_2 & Q \\
\hline
(1,15,\bar3) & (3,2) & 1 & -1 & +1/3 \\
             & (6,1) & 1 & 4 & -2/3 \\
(1,\bar {15}, 3) & (3,2) & -1 & +1 & -1/3 \\ 
               & (6,1) & -1 & -4 & +2/3 \\
(1, 24, 1) & (1,1) & 0 & 0 & 0 \\
           & (1,3) & 0 & 0 & 0 \\
           & (8,1) & 0 & 0 & 0 \\
           & (3,2) & 0 & 5 & -1 \\
           & (\bar 3, 2) & 0 & -5 & +1 \\
\end{array}
\end{equation}
Plus an extra content
\begin{equation}
\begin{array}{llrrr}
& & Y_1 & Y_2 & Q \\
(1,15,\bar3) & (1,3) & 1 & -6 & +4/3 \\
\end{array}
\end{equation}

We can arrive to the same result by chaining some
branchings. A straightforward way is $SO(32) \supset SU(16)\times U(1)$,
\begin{equation}
   \label{intoSU16}
    496= 1_0 + 120_4+ \bar{120}_{-4} + 255_0
\end{equation}

and then $SU(16) \supset SU(15)\times U(1)$ and $SU(15)\supset SU(5) \times SU(3)$,
to finish applying (\ref{15}),(\ref{24}). In this way the quarks come from the initial
 $120$s, while the leptons are from the $255$. Or respectively in $SU(15)$,
 from the $105$s and the $224$.
\begin{figure}
    \centerline{
    \includegraphics[width=1.8\linewidth]{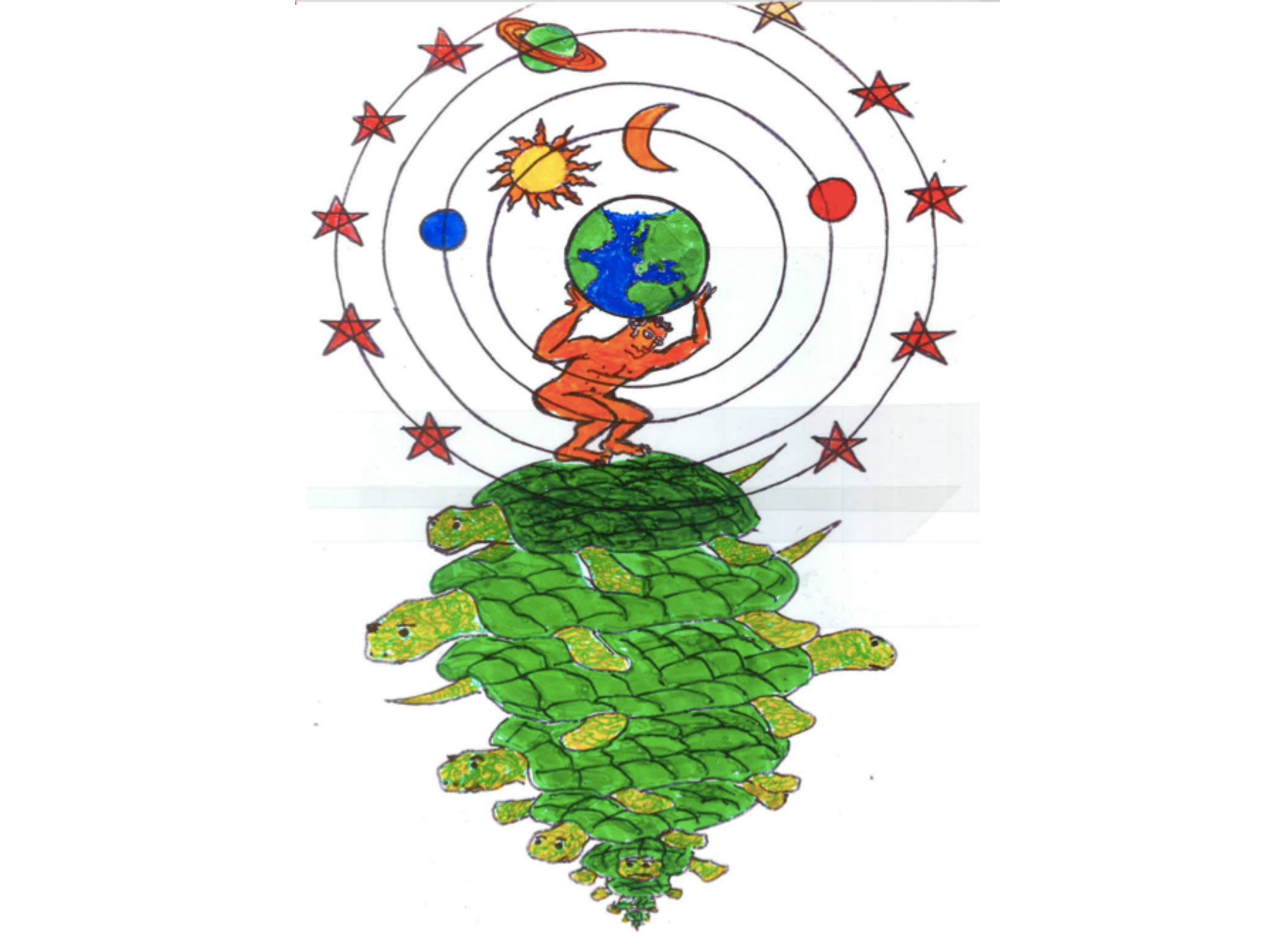}}
    \caption{Illustration of the concept \emph{Turtles all way down}, with
    an spectator but massive \emph{giant}
    (Credit of the drawing: De Rújula)}
    \label{fig:tortugas}
\end{figure}
\section{$SO(32)$ from postulates}
\label{fromPostulates}

Once we know that our aim is to get not the fermions but just
the scalar partners of a Susy Standard Model, we can wonder if
there is some set of postulates that isolates directly
the flavour group, or at least the number of generations it has.
It turns out, there is an amusing set of requirements that
force this result. 

The clue is the "recursive" property of colour:
we can get the $\bf 3$ colour triplet out
of $\bf \bar 3 \times \bar 3 = 3 + 6$.
And also we can get singlets, 
from $\bf 3 \times \bar 3 = 1 + 8$.

And adding to this hint, we notice that one quark
with an antiquark allows to build particles of electrical charges +1, 0, and -1,
but not only that: also
we can build a charge +2/3 with two antiquarks of
down type, and a charge -1/3 with one antiquark down
plus other antiquark down. This was in fact the spirit
of the above decomposition of $SU(5)$ flavour, but it 
is even more interesting when starting from particles
and going later to groups.

\subsection{Turtles and elephants}

We consider scalars as composites either of pairs of quarks, as a 
colour triplet,
or of pairs quark anti-quark, as a singlet. Furthermore, we
divide the quarks in two classes: \emph{turtles} and \emph{elephants}\footnote{Or \emph{giants}, see figure \ref{fig:tortugas}},
and add a rule: only \emph{turtles} can combine into composites.

We assume there are $N$ up-type quarks, of these $k_u$ \emph{turtles}, and
$N$ down-type quarks, of which $k_d$ \emph{turtles}.

We ask for what values of $N, k_u, k_d$ the number of scalars
of each type is exactly $2N$, as required in supersymmetry models.
This gives two equations for squarks up and down:
\begin{eqnarray}
 2N = & k_u k_d\\
 2N = & k_d (k_d+1)/2  
\end{eqnarray}
So $N\geq 3$ (actually, $N$ must be half of an hexagonal number) and $k_d=2 k_u - 1 $.
If we add other two conditions, for sleptons charged and neutral
\begin{eqnarray}
 2N = & k_u k_d\\
 4N = & k_u^2 + k_d^2 -1 
\end{eqnarray}
then the solution is unique, $N=3$, $k_u=2$, $k_d=3$. 
There are five \emph{turtles} and one \emph{elephant}, 
that we can name as the top quark.

However, note that if we consider all the combinations of
\emph{turtles} we find that we get three extra "squarks" of charge $+4/3$,
and their opposites.

\subsection{Colourless and coloured flavour groups}
The extra "squarks" look as a penalisation but group
theoretically they
are the ones that allow to complete the flavour
supermultiplet into a $15$ of $SU(5)$

At this level and without colour, we could consider that
the flavour is organized in the $54$ of $SO(10)$,
and then break it down to $SU(5) \times U(1)$
$$
54 = {15}_4 + \bar{15}_{-4} + {24}_0
$$
where again the hypercharge from this U(1) can be combined
with the one on \ref{15},\ref{24} to reproduce
the electric charge. 

If we want to incorporate colour and unify colour-flavour,
our minimal candidate is $SU(15)$. From here we 
can go up to $SO(30)$ and then to $SO(32)$ adding singlets,
or substituting colour $SU(3)$ by $U(3)$.

\section {A case for one generation}

The above argument assumed the SM and particularly
that the turtles were allowed to bind only if
they had colour. But the final particle content
is very reminiscent of the Georgi-Glashow model.

Consider the usual formulation of the model
\begin{eqnarray}
5 =& (1,2)_{-3} + (3,1)_2 \label{5}  \\
10 =&  (1,1)_{-6} + (\bar 3,1)_4 + (3,2)_{-1}
\label{10}
\end{eqnarray}
and use $Q=T_3 - Y/6$ to study the electric charge of
the colour singlets and colour triplets in each representation
of $SU(5)$

\begin{table}[h!]
    \begin{tabular}{r|cccc}
        repr & singlets & triplets & antitriplets & other \\
        1 & 0 &  &  &\\
        5 & 0, +1  & -1/3 &  & \\
\textbf{24} & \textbf{0,+1,0,-1}  & -4/3, \textbf{-1/3} & +4/3,\textbf{+1/3} & \textbf{$(8)_{q=0}$}\\
    \textbf{10} & \textbf{+1} & \textbf{+2/3, -1/3} & \textbf{-2/3} &\\
    $\bar {\mathbf{10}}$ & \textbf{-1} & \textbf{+2/3} & \textbf{-2/3, +1/3} &\\
    
        15 & +2,+1,0 & +2/3, -1/3 & & $(6)_{-2/3}$\\
        $\bar{15}$ & -2,-1,0 & & -2/3, +1/3  & $(\bar 6)_{+2/3}$\\
    \end{tabular}
\end{table}

It can be argued that the union of $10 + \bar {10} + 24 $
is the one generation version of our previous construction, albeit
with only a left neutrino. See how the $X$ boson of the
GUT model is here just the q=+4/3 scalar, and how we have
the condition of two states for each fermion. In this
case our turtles, binding, are all the members of $5$ and the
elephants, not-binding, are the members of $10$. They bypass 
the previous uniqueness proof because we are allowing 
leptons to join the game.

It could be interesting to consider variants of this
game, such as flipped $SU(5)$ \cite{Derendinger:1983aj} with some rules for the
$U(1)_X$ charges of the $24$ and $15$, or even
an anomalous "deflipped $SU(5)$" with the up quark in the
fundamental but the standard hypercharge assignments,
so that the $+4/3$ triplet would appear in the decuplet.
The authors in \cite{Buchmuller:1982tf} use a different
charge assignment to produce the +2/3 charge in the $24$,
so that it is composed of gluons, a photon, and a whole
set of $SU(3)\times U(1)$ charges, and then a broken susy $SU(5)$
produces the standard model particles as Goldstone fermions.

The $SU(5)$ model was frequently used as foundational
model in the peak of composite theories in the early eighties.
In most cases the components in the $5$ were new particles,
but some models considered to keep right fermion as elementary
and only left as composite, and even there was some proposal \cite{Fajfer:1988xa}
were the $5$ had a shared mix of preons and known fermions. The
author in \cite{Anselm:1981bi} considers a fundamental "quint" and
a composite $10$, albeit from the product of three "quints".
Early literature also includes ideas where only the leptons are
composite, as well as proposals where only the first family is elementary.

\section{The role of the top}

The argument in section \ref{fromPostulates} tells us that
there is one quark that does not act as a preon for the susy scalars. It does not tell
us which one. We need to identify the elephant. And
really we have not a concrete argument.

We favour the top quark due to the horizontal, flavour-like, symmetries
found in the first section: $SU(3)_{f_3} \times SU(2)_{f_2}$. When
considering the values of quark Yukawa couplings, and thus
quark masses, it seems more fitting that the $f_2$ symmetry
relates $u$ and $c$, with the $t$ quark being the excluded
one. This scenario is typical when breaking textures.

We were in fact inspired by the empirical observation that
toponium doesn't exist, but this is perfectly justified by
the mass of the top being closer to the Fermi scale than to the
QCD scale. So it disintegrates faster than it binds. However, 
we have not discussed the binding mechanism here, and
the one from QCD doesn't apply.

The heavy mass of the top quark can be used as an argument
in compositions where the quarks act as charges at the ends
of a relativistic open string, as they need to be massless, or at least as
light as possible.  And ultimately it can be expected that masses in 
the standard model -with right neutrinos but not other particles-
are protected by two dual symmetries: one that fixes all the degrees of
freedom as Dirac massless except the top quark, and another one that
fixes all the degrees of freedom as Majorana massless except the neutrinos
\footnote {Such protection splits degrees of freedom as 84+12 in both cases, and
thus we should look for some group representations in the $84$, $42$ 
or $21$. Note M2-brane and M5-brane carry a $SU(9)$ symmetry }. 

\section{Discussion on Masses}

The main point of this section is the existence of a mass
formula that produces pairs of equal mass particles, as
required by unbroken susy.

Excepting the top quark, preon models from the eighties were known to produce
realistic masses. One model from that age is \cite{Koide:1982si},
that became popular later due to the accuracy in the lepton
sector. It assumes that all the mass comes from a single
abelian charge and that all the preons are presented
in the same state, so the energy of a pair, having
the same spatial wavefunction, is simply the energy
of a single element with the sum of the charges 
\begin{equation}
    E(q_a,q_b)= E(q_a+q_b) =  (q_a+q_b)^2 K_\Omega
\end{equation}
Intuitively one could imagine classical charge on two 
spherical surfaces of radius $\Lambda$ and common center,
and sum both self-energies plus the interaction energy and
see how it does not depend of the radius $\Lambda$.

Furthermore, the model \cite{Koide:1982si} makes two provisions for
composites of three pairs $(a,b^i)$ to produce realistic
masses:
\begin{itemize}
    \item That the charge of the three $b^i$ particles add to zero 
    \begin{equation}
        z_1 + z_2 + z_3 = 0
    \end{equation}
    \item That the self energy of the common preon $a$ averages
    the self energy of the other three
    \begin{equation}
        z_0^2 = { z_1^2+z_2^2+z_3^2 \over 3}
    \end{equation}
\end{itemize}

The first condition is easily met extracting an abelian
charge from any direction of a $SU(3)$ triplet, and we have some.
The second is a sort of trace condition but it is imposed 
ad-hoc. Remember also that $z_i^2= (z_j+z_k)^2$. So at least some
scalars keep having the same self-energy than a third preon in the set.

Lets parametrize with an angle $\alpha$ all the possible ways to
produce an abelian charge from the $3$ representation of SU(3), in
$T_3,T_8$ basis
\begin{eqnarray}
        z_1=&\frac 12 \cos \alpha &+ \frac 1{2\sqrt{3}} \sin \alpha\\
        z_2=&0 &- \frac 1{\sqrt{3}} \sin \alpha \\
        z_3=& - \frac 12 \cos \alpha &+ \frac 1{2\sqrt{3}} \sin \alpha \\
        z_0=&  \pm 1/\sqrt 6 &
\end{eqnarray}
This is $T_3$ for $\alpha=0$ and $T_8$ for $\alpha=\pi/2$

The solutions have an obvious periodicity $2 \pi /3$ and symmetries at
$\pi/6$ and $\pi/2$. 

As it is well known, if we use a scale factor $k=m_e+m_\mu+m_\tau$ then
for $\alpha=0.745821$ the mass triplet $[k (z_0+z_i)^2]$ recovers exactly
the values $m_e,m_\mu,m_\tau$. This is sometimes interpreted as 
a prediction for $m_\tau$ \cite{Koide:1982si}, as we can use $m_e$ and
$m_\mu$ to recover $\alpha,k$ and then calculate the extant mass, well within
one sigma of the current measurement error. Note also that $k z_0^2 = 313.85$ MeV,
a familiar quantity from QCD \cite{Scadron:2007ju}.

More interestingly, we can ask for
the mass values of the octet and sextet. Note that they do not
depend on $z_0$; our choosing of sign is translated to the whole
$[z_1,z_2,z_3]$ tuple, but this is in turn just a exchange of
preons with antipreons.

\begin{table}
    \centering
    \begin{tabular}{r|ccccc}
         & 0 & $\pi/2$ & $\pi/4$ & $\alpha_{SM}$ &$\alpha_{.68583}$\\
         \hline
       $\bar u d$  & 1553.4 &914.63  & 1756.96 & 1776.9 & 1801.22 \\
 (3,2) $\bar u s$  & 313.85 &53.848  & 0 & 0.5110& 3.4179\\
       $\bar u b$  & 15.853 &914.63  & 126.144 & 105.65& 78.463\\
        \hline 
       $(\bar 3,2)$ & \multicolumn{4}{c}{\tiny{(as above})}& \\
       \hline
       $\bar s d, \bar d s$ & 470.775& 1412.32 & 1756.96 &   1717.2& 1647.7\\
 (8,1) $\bar d b, \bar b d$  & 470.775& 1412.32  & 126.144 &  91.47& 49.128\\
       $\bar b s ,\bar s b$  & 1883.1 & 0 & 941.55 &  1016.0& 1127.8\\
       $(\bar s s,\bar d d,\bar bb)$ & \multicolumn{4}{c}{0}& \\
       $\bar u u,\bar c c$ & \multicolumn{4}{c}{0} \\
        $(\bar u c,\bar c u)$   & \multicolumn{4}{c}{0 if $u = c$,  1255.4 if $u \neq c$  } \\
    \end{tabular}

    \caption{"slepton" masses for $kz_0^2=$313.85 MeV. There are
    always two massless sneutrinos in the octet (8,1) and other two 
    in (1,3). The other two combinations of (1,3) and (1,1) can be
    chosen to have zero mass too.}
    \label{tab:uncoloured}
\end{table}

\begin{table}
    \centering
    \begin{tabular}{r|ccccc}
     & 0 & $\pi/2$ & $\pi/4$ & $\alpha_{SM}$ & $\alpha_{.68583}$ \\
     \hline
        $ud$ & 15.853 & 26.93 & 42.048 & 45.18& 49.128\\
  (3,2) $us$ & 313.85 & 1829.25 & 1255.4 & 1205.3& 1127.8\\
        $ub$ & 1553.4 & 26.93 & 585.65 & 632.66& 706.162\\
      \hline
        $dd$ & 1883.1 & 627.7 & 2342.61 & 2388.9& 2445.29\\
        $ds$&470.775  & 156.93 & 42.048  & 55.31& 78.463\\
  (6,1) $ss$ &  0 & 2510.8 & 1255.4 & 1156.1& 1007.05\\
        $bd$ & 0  &  627.7 & 313.85 & 289.03& 251.76\\
        $bs$ & 470.775 & 156.93 & 585.65 & 597.21& 611.32\\
        $bb$ & 1883.1 &627.7  & 168.192 & 221.23& 313.85\\
    \end{tabular}
    \caption{"squark" masses for $kz_0^2=$313.85 MeV. Note that
    a global change of sign in $s,b,c$ exchanges the $(3,2)$ of sleptons
    and squarks}
    \label{tab:coloured}
\end{table}

\subsection{Paired scalars in the same
representation}

Our goal is to check if we can realistically recover a
mass spectrum similar to supersymmetry scalars. This means
that we should find pairs of scalars in the same 
representation having equal masses. We have 
to cases.

For $\alpha = \pi/2 $ or $\pi/6$ we recover 
two equal masses and a different one in the triplet of $(3,2)$
$$
  \frac 1{4} \pm \frac {\sqrt{2}}6, \frac 1{2}\mp \frac{\sqrt{2}}3,   \frac 1{4} \pm \frac {\sqrt{2}}6
$$
The octet is never a problem as we have the "antiparticles" in the 
same set. The sextet fails for one pair. The triplets of $(3,2)$ are
also a source of trouble, as the pair should be expected to happen
across the $SU(2)$; the only possibility is to assign the same charge to "preons" $c$ and $u$.

A better result happens for $\alpha = 0$. Here a change of sign just exchanges
charges $z_1$ and $z_3$,  so we get exact pairs of masses in all cases.
More important, also the sextet is grouped into pairs.

So we can conclude that it is possible, with this mass formula, to obtain
an spectrum that resembles the scalars of a mildly broken supersymmetry.

If we take seriously that the mass of the fermions is just
the "preonic" self energy, then the discrepancy in
the $\alpha=0$ case contains a sort of isospin 
exchange: we have assigned \begin{equation*}
  \begin{gathered}u=c=313.85, \\s=0,d=b=470.8 \end{gathered}
\end{equation*}
and we have got 
\begin{equation*}
  \begin{gathered}(u,c,t)=(0,470.8,1883), \\
  (d,s,b)=(15.8,313.85,1553.4)
  \end{gathered}
\end{equation*}

\subsection{Solutions with paired scalars in different
representation. Missing symmetry.}

Lets go a bit beyond this the scope of this work, to look also for
rotations that create some pairs in different
representations. The motivation here is to explore
for some extra symmetry that includes
quarks of different charges. This is phenomenologically
motivated by two approximate observations:
\begin{itemize}
    \item that the top mass still can fit in a Koide SU(3)
    tuple but only respect to high masses, for instance $(m_c,m_b,m_t)$.
    \item that $(m_s,m_c,m_b)$ also look as a valid tuple.  
\end{itemize}
Both observations need to get values from different 
representations. Looking for such solutions is beyond
the scope of this work, but we can cover here the cases where
a mass appears in two representations, as it is
really a completion of our inspection. 

We find two cases. First, for $\alpha=\pi/4=0.7854...$ we obtain
for the triplet of the sleptonic $(3,2)$:
\begin{equation}
 \frac 1{4} (2+\sqrt{3}), 0,  \frac 1{4} (2-\sqrt{3})
 \label{lept}
\end{equation}
and with change of sign:
\begin{equation}
 \frac 1{12} (2-\sqrt{3}), \frac 2{3},  \frac 1{12} (2+\sqrt{3})
 \label{quar}
\end{equation}

The second case happens when $z_0=\pm 2 z_i$ for some index $i$. All the
solutions are similar obtained as reflections and translations
of $\sin^{-1} {1\over 2 \sqrt 2}$ ; we show in the tables
 $\alpha=0.6858...$ for continuity.

The proportion (\ref{lept}) was first found by \cite{Harari:1978yi}
who assigned them to $(m_s, m_u, m_d)$, back in 1978.

It is interesting that if we scale (\ref{lept},\ref{quar})
to get equal masses in both tuples,
we observe a relation  $\sum m_q = 3 \sum m_l$, and that works empirically for
$(m_s,m_c,m_b)$ with respect to $(m_e,m_\mu,m_\tau)$. Also a 
comparison of table \ref{tab:coloured} respect to the
quantities of table \ref{tab:uncoloured} seems to hint 
a missed factor of three in other cases: $627.7 \to 1883.1$,
$156.93 \to 470.7$, $313.85 \to 841.55$. With
the factor three the tuple we have called $(ud, us, ub)$
gets more realistic masses and we know from elsewhere
that it can be further rotated to produce more exact values,
but we will not pursue this here \footnote{
It is just an empirical observation, that solving
\begin{equation*}
m_3 = \left( \left( \sqrt{m_1} + \sqrt{m_2} \right) \left( 2 - \sqrt{3 + 6 \frac{\sqrt{m_1 m_2}}{\left( \sqrt{m_1} + \sqrt{m_2} \right)^2}} \right) \right)^2
\end{equation*} with input (172.4, 4.183) gets 1.3495, and
then input (4.183,1.3495) gets  0.092
}

Our conclusion in this section is that the breaking down
to a single $SU(3)$ flavour, and particularly the neglect
of weak isospin, misses some extra symmetry that would
allow more comprehensive mass formulae.

\section{Discussion on related groups}
\subsection{On $SU(15)$}


For the group decomposition, similar results could be obtained 
with only $SO(30)$ or $SU(15)$ as a coloured
flavour group, or $SO(10)$ or $SU(5)$ as colourless
flavour groups, or even with ${Usp}(32)$. 

$SU(15)$ was considered as a GUT group by \cite{Adler:1989nn}
and \cite{Frampton:1989fu}. The first reference notes
that it is a subgroup of $SO(32)$ Both references embed
a full generation $$
(l_L,l^c_L,\nu_L,u_{rgb,L},u_{rgb,L}^c,d_{rgb,L},d_{rgb,L}^c)
$$
inside the fundamental of $SU(15)$. On the other hand,
our approach embeds the $(2,1)+(1,3)$ \emph{turtles} of our $SU(5)$ flavour:
$$
(u_{rgb},c_{rgb},d_{rgb},s_{rgb},b_{rgb})
$$
and we use, as noted above, the $105, \bar{105}$ and $224$ representations.

Recently \cite{Coriano:2022yzv,Coriano:2023dfu}  have considered $SU(15)$ in the context of the
standard model extended with bifermions, so they naturally use
these representations. They consider the particles to be elementary,
so "biquarks" instead of "diquarks" or mesons, but this distinction blurs away
when we consider an interpretation as open string terminated in 
quark labels. More importantly, they still keep having leptons in the
fundamental representation, so it is possible to get a lepton number
in the $15 \times 15$ and $15 \times \bar{15}$ products. 

The difference with our model is due to option for the 
breaking path $SU(15) \supset SU(12)\times SU(3)_l \times U(1)
\supset SU(6)_L \times SU(6)_R \times SU(3)_l \times U(1)\times U(1) $,
that allows to put a whole generation of the SM without right neutrinos
in the decomposition of the $15$, at the cost of some
delicate surgery \cite{Adler:1989nn,Frampton:1989fu}. 
The first extracted $SU(3)_l$ group has the goal of joining
all the leptons of each generation in a single multiplet; if we
want an extra $\nu_L^c$ neutrino it must be expanded to $SU(4)_l$
and then the whole group to $SU(16)$

\subsection{On $SU(8)$}
This section and the next one are explorative work, the
main theme being if representations of 
other groups from supergravity and string theory 
can benefit of a similar interpretation as scalars of 
some supersymmetric standard model.

$SU(8)$ appears directly because an alternate chain down from $SO(32)$ is to take the 
detour  $SU(16) \supset SO(16) \supset SU(8) \times U(1)$ 
\begin{equation}
\begin{aligned}
496  &=& 1 + 120_4 + 120_4 + 120_0 + 135_0& \\
     &=& 1  + 3(1_0 + 28_2 + \bar 28_{-2} + 63_0)& + \\
     & & +36_2+ \bar {36}_{-2} + 63_0&
 \end{aligned}
\end{equation}
And then we can go for the group theory of $SU(8) \supset SU(5) \otimes SU(3) \otimes U(1)$ but
with a lot more of hypercharge assignments (usually uglier, but worth a glance).

Family GUT unification with $SU(8)$ was examined with some detail
in 1980, see for instance the references in the recent revisit of \cite{Adler:2014pga}.
Typically three families of standard model fermions were expected to be
in the summed complex representation $\bar 8 + \bar 28 + 56$
and some criteria was used to select the hypercharge assignments.Most
models preferred to interpret for flavour the 
first $SU(3)$ in $SU(8) \supset SU(5) \otimes SU(3) \otimes U(1)$ 
instead of leaving it for colour as \cite{Gell-Mann:1976xin}.
Both approaches differ only in the algebra of $U(1)$ charges for the multiplets. The fundamental decomposes as 
a colour triplet, a $SU(2)$ horizontal doublet,
and a $SU(3)$ horizontal triplet.
$$
8=(1,1,3)_{0,-5} + (1,2,1)_{-3,3} + (3,1,1)_{2,3}
$$
Note it went first to 
$$
8 = (1,3)_{-5} + (5,1)_3
$$
and while in the first approach $SU(5)$ is flavour-colour,
in the second it is just two horizontal symmetries
and the colour triplet is explicit. So we prefer this later
way because so all the irreducible representations
of $SU(8)$ have an interesting interpretable descent. The decomposition of the $28$ 
has a quark content, triplets, that looks very much as
our division in five turtles and
one elephant, 
$$
28 = (1,\bar 3)_{-10} + (5,3)_{-2} + (10,1)_6
$$
but it is different to the $SO(32)$ case. To illustrate a particular assignment, if we think of 
the fundamental as "half-charged preons" of charges $\pm 1/2$, $1/6$, then in the $28$:
\begin{itemize}
    \item[-] $(1,\bar 3)$ is one anticoloured particle of charge $+1/3$
    \item[-] $(5, 3)$ are coloured particles, three of charge -1/3, two of charge $+2/3$
    \item[-] $(10,1)$ contains six particles of charge 0, three of
charge $-1$ in an horizontal "antitriplet"\dots and one of charge $+1$
\end{itemize}

So this content doesn't allow for our "recursive" interpretation of the interplay
between the $32$ and the $496$ of $SO(32)$

We can play also with content from extra representations. The $36$ somehow complements the $28$, and
the $63$ can provide a full uncoloured $(24,1)$ to break into
different charges. Besides, in this path, the fundamental of $SO(32)$ appears in $SU(8)$ as
\begin{equation}
\label{the32}
32 = (8_{1,2} + \bar 8_{-1,2})+(8_{1,-2} + \bar 8_{-1,-2})
\end{equation}
and so it provides extra $U(1)$ charges and extra particles;
one needs a good motivation to justify a particular pick. We can explore one hundred 
weightings to extract the electric charge $Q$ of each representation, most of the combinations offering
extra quark and lepton content, including some $+4/3$ quarks.


We could also use the process via via $SU(5) \supset SU(2) \otimes SU(3) $ to assign weak and colour
multiplets as usual. On our point of view, both $SU(2)$ and $SU(3)$ here are horizontal groups. 

One can observe that (\ref{the32}) meets the condition
asked in \cite{Gell-Mann:1976xin} of having only singlets and triplets of colour, and so wonder what reasons, besides
simplicity, motivate the exclusion from the listing.

We could also consider first a regular descent, via $SO(16)$ to
$SU(8) \times SU(8)$
$$ 120 = (8, 8)_0 + (28, 1)_2 + (1, 28)_{-2} $$
$$  255 = (1, 1)_0  + (8, \bar 8)_2 + (\bar 8, 8)_{-2} 
+ (63, 1)_0 + (1, 63)_0$$

\subsection{ On $E_8 \times E_8$}

Exotic approaches to flavour
are not unknown in supergravity, a good example being the diagonal $SU(3)$
from Gell-Mann, that also ignores electroweak 
charge \cite{Nicolai:1985hs}. And as $SO(32)$ is relevant
to 10D sugra, and all the 
10D supersymmetric theories are related via string dualities, it is interesting
to speculate if other corner of this
space, the $E_8 \otimes E_8$ group, 
can present a similar mix. 

We can examine this possibility
starting from the conclusions of the above sections,
albeit at the moment the discussion will be very light,
and  inconclusive, if not disappointing.

$E_8$ is not considered in \cite{Gell-Mann:1976xin} because
the authors apply a "colour restriction" in their selection of groups,
asking for decomposition of the fundamental representation having
only singlets and triplets of $SU(3)$. It is
more particularly reviewed by \cite{Adler:2002yg}, who
enumerates the problems to use it as a group GUT and also 
considers decomposition with explicit family group $SU(3)_F$.
A separate approach with explicit colour group $SU(3)_c$ and
then mixed electroweak-flavour $SU(6) \times U(1)$
was done in \cite{Baaklini:1980uq} via an initial breaking into $SU(9)$.
Generically, $E_8$ has an industry of 
its own for pure algebraic approaches, linked to Clifford algebras,
and full of interesting observations,
but reviewing it is out of the scope of this letter.

Both $SO(32)$ and $E_8 \otimes E_8$ have a subgroup $SO(16) \otimes SO(16)$.
The branching of $SO(32)$ to this subgroup is
$$496 = (120, 1) \oplus (1, 120) \oplus (16, 16)$$
very similar to the branching we have used in (\ref{intoSU16})

Isolately, each $E_8$ branches to $SO(16)$ as
$$248 = (120) \oplus (128')$$

What we suspect is that quark and lepton parts have different
roles, the quark part coming from $120$; one of the $120$s will
provide the quark-like charges, the other will provide the
antiquark ones. The lepton part can be extracted from 
the $28$ of $SU(8)$ but it could also come from the
$63$, and then we should investigate the $(128')$ irrep.

Remember that in the initial sections the critical part has been
to obtain a $15$ representation of $SU(5)$ associated
to a triplet $3$ of $SU(3)$, as well as a
$24$ associated to the singlet. And here is the problem:
any further factorisation of $SO(16)$ fails to
get representations as big as the $15$. We
are down to fives and tens too soon. Amusingly, we 
could also consider a directly branching $E8 \supset SU(5) \otimes SU(5)$; this is exploited in model building, for instance \cite{Baaklini:1980fv,Adler:2014pga}, 
but with different assignments to colour and flavour. If we
use this kind of decomposition and we accept the irreps $5$ and $10$ instead
of the $15$, it amounts to exchange some of the $\pm 4/3$
and $\pm 2/3$ charges by an excess of $\pm 1/3$ charges. 
 

\section{Discussion}

The postulate \emph{It is turtles all the way down}\footnote{I first heard
this idiom in a talk from Alvaro de Rujula in 1986} applied
solely to squarks already fixes the number of generations to be greater or 
equal than three. Adding a reasonable condition on the building of 
neutral sleptons, it fixes uniquely $N=3$ and then also the separation
between five light quarks and one heavy one that does not participate
in the composites. Of course this uniqueness is not seen when going directly
from  the $SO(32)$ group down to flavour times colour, but even in this
case there is a separation between five "turtles" in the fundamental
of $SO(32)$ and a non-participant "elephant".

While eventually all the extant multiplets of the decomposition
should be explained, the $(1, 3)$ squarks, of charge $\pm 4/3$,
are specially puzzling. They can not be organised as three 
generations of partners of four-component Dirac 
quarks. Still, they have a role in
the flavour multiplet, and they could exhibit their singularity
if chirality is introduced back in the game. We have
not considered other flipped descents that could result in
different extra $squarks$

The symmetry between quarks and diquarks or its hadronic
equivalent is known to be one of the historical origins 
of supersymmetry \citep{Miyazawa:1968zz,GaoHo} and it is used in hadronic phenomenology. But a concrete hadronic construction of
our scalars as real diquarks produces the ones of odd parity, that are excluded of phenomenological discussions as
they do not survive the 'single mode approximation' \citep{Jaffe:2004ph}.
Thus the composite "squarks" and "sleptons" bound here should be not
the ones, diquarks and mesons, found at QCD scale, but it is intriguing that
they are similar in number and mass.

Approximate supersymmetry between quarks and diquarks, to be used for dynamical
supersymmetry between barions and mesons, is a well known technique, see 
the review \cite{Anselmino:1992vg}. One of the authors of this review extended
this technique to preons \cite{Dugne:1998hk}, so that dipreons are susy partners 
of preons, but the model does not produces a dynamical susy
with new scalars as partners of leptons and quarks.

We can justify the uplift from $SU(15)$ to $SO(32)$ by
asking particle colour to be in a slightly greater group, such as $U(3)$. This
could be a hint of the difference between the binding mechanism needed here,
that should happen at high energy scale, and the usual binding of mesons and diquarks.
Observe that the usual binding shares some properties:
the top quark, our \emph{elephant}, does not bind into mesons -because it disintegrates
before-, and the masses of mesons and diquarks are in the same range of
energies that the SM fermions, as expected of a slightly broken supersymmetry.

While the composites point to HC, TC or ETC models, one must recognise that 
the motivation for $SO(32)$ is not only to produce one hypercharge
and the adequate multiplets in the decomposition, but also because of
its role in string theory. The postulates of composition need a pairing
that looks similar to labels in terminated open strings. The composition 
process from the point of view of a terminated "QCD string"
bears some similarity to the techniques 
of \cite{Armoni:2003fb} using "planar orientifolds".

 If we get the scalars from $SO(32)$ a natural question
 is where the superpartners -the actual fermions- are. They
 could be obtained by applying the susy transformation.
 In string-inspired GUTs, they should be in usual
 non-susy models for three generations, and the issue
 would be to reassure the compatibility with the $SO(32)$ group.
 In preon constructions, our focus on scalars, 
 and thus in pairs of fermions, makes the results to differ
 from most previous approaches. 
 To recover a fermion, one must consider
 an extra object, particle or string, providing again an 1/2 spin.

\section{Conclusions}
In conclusion, lets review what we have got. 
 We offer a novel interpretation of the $SO(32)$ group within the context of supersymmetric models, 
 emphasizing its potential as a flavour group for scalars. The decomposition
 and hypercharge assignment that allows to recover three generations
 has not been presented in the literature explicitly. This is for the
 obvious reason that it recovers scalars, not fermions. But on
 the other hand, to look for scalars avoids to address the 
 problem of the lack of chiral fermions.

 Besides, we offer a composite explanation for scalars of the SSM, that
 fixes the number of generations and limits the possible groups
 that can be used to generate flavour with a separate colour factor.
 In the list of possible groups, $SO(32)$ stands up. 
 
 Our postulate
 is, certainly, exotic: it suggests that while SSM fermions could
 be elementary, the SSM scalars are composites, with
 their preons being a subset of the observed SM fermions. Far fetched as this
 postulate looks, it reproduces the $SO(32)$ decomposition
 and fixes the number of possible generations.  Also, it 
 justifies the non observation of superpartners;
 supersymmetry could be hiding in plain sight, not broken but distorted.

 We have recalled a classical mass formula that when applied to
 our piece of $SU(3)$ flavour symmetry can grant pairs of scalars
 having the same mass. 
 
 The decomposition seems to imply that each generation also includes two
 extra "scalar quarks" of charge $\pm 4/3$. It is unclear if 
 such scalars could have an associated fermion, as 
 it should be of Weyl type, not Dirac. On other hand, our approach
 could be compared to the "scalar democracy" of \cite{Hill:2019ldq} that
 goes further and proposes the existence of a scalar bound state 
 for every pair of fundamental fermions, either leptons or quarks.

\section{Acknowledgements}
This work was supported in part by project No. PID 2022-136374NB-C22, funded by the Spanish MCIN/AEI and by the European Union. 
Figure (\ref{fig:tortugas}) adapted by Álvaro de Rújula from his personal archive. I also thank M Porter for discussions.

\end{document}